\documentclass[12 pt]{article}
\usepackage{multicol}

\renewcommand{\baselinestretch}{1.} % Uncomment for 1.0 spacing between lines

\begin{document}

\thispagestyle{empty}

{\large
\parskip 10 pt

\hfill{}

\vspace{1.5 in}
\centerline{\LARGE\bf Magnetic Fields in Stellar Astrophysics}
\vspace{1cm}
\centerline{\bf A White Paper Submitted to the Astro-2010 Decadal Survey}

\vspace{1.cm} 
\hfill{}\\
\vspace{5 pt} 
Dmitri A.~Uzdensky, Princeton University; \ uzdensky@astro.princeton.edu \\
\vspace{5 pt} 
Cary Forest, \ University of Wisconsin, Madison \\
\vspace{5 pt} 
Hantao Ji, Princeton Plasma Physics Laboratory \\
\vspace{5 pt} 
Richard Townsend, \ University of Wisconsin, Madison \\
\vspace{5 pt} 
Masaaki Yamada, Princeton Plasma Physics Laboratory

\vspace{1cm} 
%{\it Endorsed by the Topical Group on Plasma Astrophysics 
%of the American Physical Society}
%
Endorsed by the Center for Magnetic Self-Organization in Laboratory \&
Astrophysical Plasmas (www.cmso.info./), an NSF Physics Frontier Center
in partnership with DoE. \\

\vspace{1 cm}
\underline{Science Frontier Panel:} 
\quad Stars and Stellar Evolution (SSE)
}
\parskip 0 pt

%\date{February 13, 2009}

%--------------------------------------------------------------------

%Stars and Stellar Evolution (SSE) SFP Description:
%
%"This panel will consider science opportunities and themes surrounding
%stars and stellar evolution, including the Sun as a star, stellar 
%astrophysics, the structure and evolution of single and multiple stars, 
%compact objects, supernovae, gamma-ray bursts, solar neutrinos, and 
%extreme physics on stellar scales."
 
%--------------------------------------------------------------------
\newpage
\setcounter{page}{1}

\centerline{\bf\Large Magnetic Fields in Stellar Astrophysics}

\vspace{15 pt}
\centerline{\bf\large D.~Uzdensky, C.~Forest, H.~Ji, R.~Townsend, \& M.~Yamada}

\vspace{10 pt}
\centerline{\large A white paper submitted to the Astro-2010 Decadal Survey}

\vspace{10 pt}
{\it Endorsed by the Center for Magnetic Self-Organization in Laboratory \&
Astrophysical Plasmas (www.cmso.info./), an NSF Physics Frontier Center
in partnership with DoE.}

%---------------------------------------------------------------------

\section{Introduction}

Building a comprehensive physical picture of stellar structure and 
evolution is one of the greatest triumphs of 20th century Astrophysics. 
However, many important aspects of the life cycle of stars are still not 
understood. They include: the origin and evolution of the stellar magnetic 
field, including the large-scale field; the sun-spot cycle; the evolution 
of the stellar rotation; the origin and character of differential rotation; 
high-energy coronal activity; mass loss in massive stars; and various 
aspects of binary evolution. 
Even more critical questions remain open regarding the end of the life
of stars and their afterlife as compact objects. These include the 
mechanisms of core-collapse supernovae (SNe) and gamma-ray bursts
(GRBs), the origin of magnetars, the workings of radio-pulsars
and pulsar-wind nebulae; and some aspects of accretion disks and 
jets in binary systems. 
All these questions are at the forefront of modern Astrophysics.
Importantly, many of them unavoidably involve magnetic fields 
interacting with the plasma. 
Therefore, they belong to the realm of {\it Plasma Astrophysics}.

%--------------------------------------------------------------------

A common theme in plasma astrophysics is 
the {\it life cycle of magnetic fields}:
How are they produced (dynamo)? How do they interact with the plasma 
(magnetic braking; magnetocentrifugal wind acceleration; MHD 
instabilities such as MRI, kink, and Parker)? And how are they 
destroyed (reconnection)?

In this white paper we outline some of the most outstanding 
questions and emerging science opportunities related to 
Magnetic Self-Organization and Plasma Astrophysics.
We stress the importance of a well-balanced research program 
involving four inter-connected components: astronomical 
observations, analytical theory, computer simulations, and
laboratory experiments.

%Astronomical science funding over the next decade should be focused on 
%appropriately supporting each of these areas in a way that maximizes 
%further key discoveries and new insights (Phil Kronberg's language). 

%-----------------------------------------------------------------

\section{The origin of magnetic fields in stellar objects} 

The origin of magnetic field in stars and compact stellar remnants
is an old outstanding problem in Astrophysics. It encompasses young 
stars, main sequence stars, magnetic white dwarfs, and neutron stars, 
including magnetars. 

%-----------------------------------------------------------------

\subsection{Magnetic Dynamo in low-mass stars}

In the Sun and other cool, low-mass stars, vigorous magnetic activity 
is believed to result from the conversion of convective and rotational 
mechanical energy into magnetic energy by MHD dynamo. 
Although the basic ideas are now well established, 
we still don't really understand many key aspects, 
such as the differential rotation, the intermittent character 
of the surface magnetic field distribution (e.g., sunspots), 
the origin of the solar cycle, the strength of the large-scale 
(dipole) magnetic field, and the role of the (differential) 
rotation in the large-scale dynamo.
Significant progress on these problems is expected in the next 
decade due to a confluence of recent advances in analytical 
theory and numerical simulations, and the advent of new laboratory 
experiments.

Of particular interest in astrophysics is the origin of large-scale
magnetic fields. The kinematic (linear) dynamo theory has been advanced
recently with the development of a statistical theory unifying large-scale 
and small-scale dynamos~\cite{Malyshkin_Boldyrev-2008}.
In addition, incorporation of magnetic helicity conservation into 
non-linear large-scale dynamo theory has resulted in the development 
of a successful saturation model that matches simulations for closed 
volume~\cite{Blackman_Field-2002, Blackman_Brandenburg-2002}, as well 
as for open volume dynamos that can overcome resistive quenching, 
sustained by helicity flux~\cite{Blackman_Field-2000, Vishniac_Cho-2001, 
Sur_etal-2007, Vishniac-2009}.

Finally, there are now several operating and proposed experiments 
utilizing {\it liquid metals} to study MHD dynamo under controlled 
and well-diagnosed laboratory conditions.
In addition, a proposed {\it plasma} dynamo experiment can offer 
better diagnostics capabilities and can extend the liquid-metal
studies to more astrophysically relevant parameters, e.g., to much 
higher magnetic Reynolds numbers than in liquid metal experiments.
Importantly, it will also allow the viscosity~$\nu$ to be varied 
independently of the resistivity~$\eta$, giving access to a wide 
range in the magnetic Prandtl number $Pr_m\equiv \nu/\eta$, thought 
to be a critical parameter for the dynamo onset (in liquid metals 
$Pr_m$ is fixed by material properties and is usually $\ll 1$). 
Although the high-$\beta$ plasmas needed for such experiments 
have never been studied in the lab, recent advances in permanent 
magnet technology, plasma source development, and new schemes for 
driving flow now make it possible.
Such an experiment would be a relatively modest investment 
compared with telescopes and space missions.

%-----------------------------------------------------------------

\subsection{Magnetic fields in massive stars}

%Stable, organized magnetic fields are known to exist in about 20 stars
%with spectral types B3 or earlier.  However, most of these objects are 
%relatively low-mass ($\sim 9 M_\odot$) B2-type stars. In contrast, because
%of various technical observational limitations, magnetism in more massive 
%hot stars has so far been poorly explored and only very few genuine high-mass 
%stars have been identified as magnetic. In fact, no systematic surveys of 
%magnetism in B and O-type stars have been yet undertaken (although a Large 
%Program is now underway at the CFHT to begin to address this issue) [REF]. 
%This is why, historically, it has been assumed that magnetic fields in OB 
%stars are very rare, and perhaps altogether absent in stars with masses 
%above 8 solar masses. 

During the past few years, high-resolution spectropolarimetric studies 
have discovered $\sim 1$~kG large-scale magnetic fields in early-type 
(OB) stars with masses up to 50~$\rm M_\odot$~\cite{Townsend_etal-2008}. 
This important discovery has fundamental implications for our 
understanding of the observed properties of hot, massive stars, 
their evolution, and their final stages such as GRB-producing 
collapsars and the magnetic properties of neutron-star remnants.
However, in contrast to magnetic fields in convective low-mass stars, 
the origin of magnetic fields in massive stars 
(above $\sim 2{\rm M_{\rm \odot}}$) remains poorly understood. 
Since these stars don't have convective envelopes where a turbulent 
dynamo may operate, one expects their magnetic fields to be very
different from those of low-mass stars. And indeed, observationally, 
they are structurally simpler, and often stronger, than the fields 
of cool stars~\cite{Wade-2003}, and their characteristics show no clear 
correlation with stellar age, mass, or rotation~\cite{Kochukhov_Bagnulo-2006}, 
suggesting a fundamentally different field origin. 
In particular, it is now believed that they may be fossil 
remnants from the stars' formation stage, instead of being 
actively generated by an internal dynamo, although this is 
still debated~\cite{Braithwaite_Nordlund-2006}. Distinguishing 
between these two alternative scenarios will be a central theme 
for the next decade, and will have a broad impact on massive-star 
astrophysics.
Determining the fields' origins will shed light on their configuration 
beneath the photosphere. Given that the internal rotation profile of 
a massive star is a critical factor for determining whether it will 
produce a GRB, (e.g., \cite{Petrovic_etal-2005}), 
it is clearly important to constrain the structure of 
the interior fields capable of redistributing angular momentum.

%There are several reasons for this dearth of information about magnetism 
%in massive stars.  First, the stellar winds of massive stars are sufficiently
%strong and turbulent that they inhibit the accumulation of strong
%chemical peculiarities (such as the He and Si peculiarities observed
%in the helium-strong stars) in the stellar photosphere. Thus strong
%chemical peculiarity cannot generally be used as a proxy for magnetism. 
%Secondly, the optical spectra of massive stars contain relatively few 
%spectral lines from which to diagnose the presence of a magnetic field.
%In addition, massive stars are systematically rapid rotators,
%generating broad spectral lines which provide decreased sensitivity to
%the presence of magnetic fields. Finally, the spectra of massive stars
%are strongly influenced by non-LTE effects produced in their
%photospheres and winds, generating complex and variable lines profiles
%that are a challenge to interpret.  For these reasons, magnetism in
%hot, massive stars has been poorly explored. 

%-----------------------------------------------------------------

For robust and sensitive field measurements, many line profiles must 
be observed simultaneously in Stokes IQUV; this necessitates the use 
of echelle spectropolarimeters, putting high demands on the light 
collecting power. Therefore, the key to future observational progress 
will likely lie in the deployment of echelle spectropolarimeters on 
the new generation of larger, more sensitive telescopes.

%Challenges faced on the theoretical side also include the need for better 
%approaches to numerical modeling of massive-star magnetospheres. These 
%systems exhibit a huge range of plasma betas, extending from unity all 
%the way down to $10^{-6}$ or less. Because such tiny betas are difficult to 
%obtain in terrestrial settings, these magnetospheres represent ideal 
%natural laboratories for driving forward our understanding of plasma 
%physics in the strong-field limit. However, they represent a stumbling 
%block for standard MHD codes, as the very fast Alfven velocities 
%(approching the speed of light) imply exceedingly short Courant timesteps. 
%Either we must assign ever-greater computing resources to the problem, 
%or for instance develop new numerical algorithms that can smoothly 
%transition between standard MHD and 'rigid-field' hydrodynamics."

%-----------------------------------------------------------------

\subsection{Neutron stars}

An important astrophysical application of the large-scale 
dynamo theory is to newly-born proto-neutron stars (PNS). 
This topic is important for understanding the magnetic-field 
distribution of neutron stars and, in particular, explaining 
why some neutron stars (magnetars) have large-scale magnetic 
fields of up to $10^{15}$~G, whereas others have much weaker 
fields ($\sim 10^{12}$~G).
Also, understanding where and when a large-scale PNS magnetic 
field is generated is needed to to access the dynamical role 
of magnetic fields in core-collapse SN and GRB explosions.

%-----------------------------------------------------------------

\section{Dynamical Role of Magnetic Fields in Accretion Disks}

Magnetic fields are not just passive tracers of the gas motions 
in various astrophysical systems. Often, they play a decisive
dynamical role that may be manifested in a number of, sometimes
subtle, ways. 
This includes situations where a large-scale magnetic field governs 
the motion of the plasma, e.g., in launching, acceleration, and collimation 
of outflows , including relativistic jets; magnetically-channeled accretion 
on neutron stars, white dwarfs, and young stars; and gradual magnetic braking 
of accretion disks and stars.
But dynamical effects of magnetic fields also include the effects 
of energetically sub-dominant small-scale turbulent magnetic fields 
in accretion disks and stars on the effective long-time-scale transport 
of some large-scale quantity, such as angular momentum or large-scale
magnetic flux. This often involves a complicated interplay between 
magnetic fields, large-scale rotation, and turbulence. Perhaps, the 
most notable recent example is the magneto-rotational instability (MRI) 
and its role in the angular momentum transport (AMT) in accretion disks.

We also would like to note that MRI is just one example of 
a general situation where dynamical activity is brought about 
by some magnetically-induced instability. Other astrophysically 
important MHD instabilities include the Parker instability and 
its role in the formation of magnetized coronae of stars and 
accretion disks, and the kink and its effect on stability and 
survival of jets. Whereas linear stability analysis is rather 
straight-forward in most cases, the long-term non-linear behavior 
of most of these instabilities is much less understood, even though 
it is usually more relevant for observations.

Importantly, whereas the main focus of accretion disk research 
in the past decade has been on the AMT problem, plasma turbulence 
in disks goes far beyond the AMT problem. The focus is shifting 
towards more refined plasma physics and towards the questions of 
energy dissipation, thermodynamics, and radiation. Indeed, MRI 
leads to conversion of gravitational/rotational energy to 
turbulent kinetic and magnetic energy, but how and where is
this energy eventually dissipated into heat and observable radiation? 
How intermittent is energy dissipation? Does it occur in the collisional
(resistive MHD) regime or in collisionless regime requiring more 
sophisticated plasma physics? Do dissipation events in the disk 
directly produce non-thermal particles? How much of the accretion 
energy escapes and is dissipated in the overlying corona, producing 
the observed high-energy emission? What is the role of coronal magnetic 
fields in the overall AMT? Also, how does the small-scale MHD turbulence 
in the disk interact with a large-scale external field of the central 
star and how is the material transferred from the disk to the star? 
All these questions will likely dominate the studies of accretion disks 
in the next decade.

Numerical studies of MRI have seen a resurgence of activity 
in recent years due to newly emergent concerns about numerical 
convergence of previous MHD simulations. The ensuing scientific 
debate has demonstrated that understanding of the nonlinear saturation 
of MRI and the associated AMT requires very high-resolution simulations, 
with Reynolds and magnetic Reynolds numbers of many thousands. 
These simulations are now finally becoming feasible, and one
should continue pushing in this direction.

Experiments aiming to realize MRI in the lab and to study 
the resulting AMT have begun in the past decade~\cite{Ji_etal-2006}. 
%As in the numerical simulations where boundary treatments 
%are under debates, many technical challenges in the experiments
%come from the boundary effects. We expect that these technical 
%challenges will be met in the coming decade and 
%
The high-quality data on the MRI saturation from these experiments
will be compared quantitatively to numerical simulations; such
one-to-one comparisons should provide much needed physical insights 
and serve as benchmarks for numerical techniques.

One further development in the MRI and dynamo areas is to 
consider collisionless plasma effects beyond single-fluid MHD
\cite{Quataert_etal-2002, Sharma_etal-2003, Schekochihin_etal-2005}.  
In collisionless plasmas, e.g., in hot accretion disks, compression 
of magnetic field lines causes the plasma pressure to become 
anisotropic, modifying the linear stability of the MRI and 
shifting the fastest-growing mode to larger scale. In addition,
the pressure anisotropy drives various kinetic micro-instabilities
(e.g., firehose and mirror), which has a profound effect on the
system.
New experiments investigating plasma flow-driven instabilities may 
be an important way to study the microphysics of collisionless~MRI, 
and several such experiments are now under way. Comparisons between 
liquid-metal and plasma experiments will elucidate these uniquely 
plasma, non-MHD effects.

%------------------------------------------------------

\subsection{GRBs}

Recently, it has be recognized that traditional, purely 
neutrino-driven models of supernova (SN) and especially 
GRB explosions may be inadequate. This has lead to a growth 
of interest in magnetically-driven core-collapse explosions 
of massive stars, including the application of the magnetic 
tower concept. Therefore, understanding classical magnetic 
processes (such as dynamo, MRI, Parker, and kink instabilities, 
magnetic reconnection, relativistic jet acceleration and collimation) 
in the exotic environment of collapsing stars is crucial.

Making progress on these issues will require sophisticated numerical 
simulations. Fortunately, several (general-)relativistic MHD codes 
incorporating the relevant microphysics (including realistic neutrino 
transport) are now being developed. These powerful codes will be capable 
of addressing the above issues and may lead to a breakthrough in GRB 
(and~SN) central engine modeling in the next few years.

In addition, there is also an intriguing possibility 
that GRB jets may remain magnetically dominated out 
to large distances~\cite{Lyutikov_Blandford-2003}. 
The emitted gamma radiation may then be powered 
by magnetic reconnection in current sheets produced 
by kink instabilities \cite{Drenkhahn_Spruit-2002, 
Giannios_Spruit-2006} or by initial irregularities in the jet.
Figuring out how to distinguish observationally this scenario 
from the classical internal shock model is an important challenge 
in plasma astrophysics for the next decade.

%-----------------------------------------------------------------

\section{Magnetic Reconnection and Coronal Activity}

Whereas magnetic fields are usually created in dense, cold regions 
by dynamo mechanism, their destruction often takes place in the 
surrounding hot and rarefied coronal regions, through the process 
of {\it magnetic reconnection}.
Reconnection is usually defined as a rapid rearrangement of the 
magnetic field topology, but its importance in astrophysics stems
from the associated a violent, explosive release of magnetic free 
energy and its conversion to the plasma kinetic, thermal, and 
non-thermal particle energy.
Reconnection is believed to power for solar/stellar flares 
and coronal heating, magnetic storms in planetary magnetospheres, 
and flares in accretion-disk systems; it has been also invoked 
as a possible mechanism for giant magnetar flares, for GRBs, and 
for solving the pulsar-wind $\sigma$ problem. It may also be an 
important source of energetic non-thermal particles (cosmic rays).
Finally, reconnection may also be crucial for the creation of 
large-scale ordered  fields.

Whereas single-fluid resistive MHD is probably a good description 
in the dense regions where magnetic fields are generated, it may 
not be applicable in the overlying tenuous plasmas where reconnection
takes place. Thus, the life of magnetic fields is not limited to MHD 
but goes well beyond, into the ``hard-core'' plasma physics.

The past decade has witnessed great progress on magnetic reconnection, 
brought about by the confluence of powerful numerical simulations,
dedicated laboratory experiments, and analytical theory.
These complementary approaches paint a compelling physical 
picture emphasizing the critical role of plasma collisionality. 
This picture can be summed up by stating that there are two 
regimes of reconnection: a slow reconnection in collisional
plasmas where resistive MHD is applicable, and a fast reconnection 
regime in collisionless plasmas; the latter can be realized by either 
the two-fluid (e.g., Hall) effects or by anomalous resistivity due to 
kinetic plasma micro-instabilities.

However, despite its wide acceptance in space and plasma physics 
communities, this new understanding of reconnection is only now 
starting to be applied in astrophysics. In particular, it helped 
elucidate the self-regulatory nature of solar/stellar coronal 
heating and lead to the formulation of the concept of marginal 
collisionality of astrophysical coronae (including accretion-disk 
coronae of black holes~\cite{Goodman_Uzdensky-2008}). 
This example illustrates that applying the results of basic plasma 
physics research, including lab experiments, can be quite fruitful 
for astrophysics.

%------------------------------------------------------------

Even with the above recent advances in the physics of reconnection, 
many important questions concerning both its fundamental aspects 
and its astrophysical applications remain unsolved. 
Here are some of the most outstanding questions that will dominate
astrophysical magnetic reconnection research in the next decade:

1. {\it Impulsiveness:} 
Reconnection often occurs impulsively.%\cite{Bhattacharjee} ? 
Is there any general criterion why magnetic energy is 
stored for a long period and then suddenly released? 
How do anomalous resistivity and/or two-fluid effects
produce an impulsive transition to fast reconnection?  
Which of them dominates in a given situation and how 
do they interact with each other?

2. {\it Energy partitioning and non-thermal particle acceleration:}
To understand various astrophysical systems such as stellar and 
accreting black-hole coronae, it is important to know where the 
magnetic energy released by reconnection goes, i.e., how it is 
partitioned among the bulk kinetic energy of the plasma, the electron 
and ion heating, and non-thermal particle acceleration.

Specifically, non-thermal particles often play an active role
in low-density coronal environments. For example, a large 
fraction of the magnetic energy released in solar flares 
is believed to go to non-thermal electrons. This raises an 
intriguing possibility that reconnection may be an important 
source of cosmic rays. And yet, the efficiency of particle 
acceleration in reconnection is still not understood.
Determining it requires examining the interplay between 
plasma microturbulence, non-thermal particle acceleration, 
and various loss mechanisms.

%These questions are very important, for example, for coronae of 
%accreting black holes. Indeed, these hot ($T_e \sim 100$~keV) 
%coronae are believed to be responsible for X-ray emission observed 
%in these systems, extending to hundreds of keV. Magnetic reconnection 
%is often invoked as the main heating mechanism for both electrons and 
%ions, and the density is low enough for the electrons and ions to stay 
%thermally decoupled from each other (with $T_i \gg T_e$).

%4. Is reconnection necessary for large-scale dynamo?
%What is the role of CMEs in evacuating excess helicity 
%in the sun and thus regulating the solar magnetic cycle?

%5. What governs the observed power-law distribution of solar flare energies?

3. {\it High-Energy-Density Reconnection:}
Several important astrophysical phenomena where reconnection 
has been invoked, namely giant flares in the magnetospheres 
of magnetars and central engines of gamma-ray bursts, have
physical conditions that are very different from those 
considered in the reconnection research so far. In particular, 
the energy density in these systems is so high that one needs 
to take into consideration pair creation, radiation pressure, 
and radiative cooling. Such high-energy-density reconnection 
has not been explored so far and represents a new frontier 
in astrophysical magnetic reconnection.

4. {\it Non-stationary, turbulent reconnection:}
Most classical reconnection models are steady-state and laminar. 
A new research frontier is time-dependent, turbulent, reconnection 
(both collisional or collisionless) in very large (especially 
astrophysical) systems whose global scale is many orders of magnitude
larger than the microphysical plasma scales.
Modern large-scale simulations (both fluid and PIC) indicate that 
the nature of reconnection in such systems is going to be profoundly 
different---they are more prone to virulent secondary instabilities, 
such as the tearing of a long current sheet, causing it to break up 
in a chain of multiple plasmoids and probably affecting the reconnection
rate scaling. This is a particular example of the interplay between 
reconnection and turbulence, which is becoming an important theme in 
reconnection research. 
It is important for astrophysics because of its relevance 
for non-thermal particle acceleration~\cite{Drake_etal-2006} 
and its observable signatures (e.g., radio emission).

%Work is now underway to investigate the dynamical effect of the plasmoids, 
%e.g., on the time-averaged reconnection rate.

%8. How are current sheets --- potential sites for reconnection -- 
%are created in various astrophysical systems. For example, in 
%the solar and accretion disk coronae, what dynamical processes 
%determine the distribution of sizes and strengths current sheets?
%It is necessary to develop models which would lead to formation 
%of a large number of reconnection layers. 
%Another unresolved issue is the effect of line-tying of the magnetic field 
%at the boundary, which is expected to affect the stability of plasmas 
%and the reconnection rate.
 
%6. One of the major questions still remaining is: How do global systems 
%generate local reconnection structures through formation of one or 
%multiple current sheets, either spontaneously or forced by boundary 
%conditions? 

%--------------------------------------------------------------------

%{\it Science opportunities}:

%As mentioned above, progress in astrophysics comes from different 
%aspects of the modern research enterprise: observations, theory, 
%simulations, and laboratory experiments.

Future progress in numerical simulations of reconnection will 
come both from the expected increases in computer power and 
from advances in numerical codes. For example, incorporating 
Coulomb collisions into a PIC code~\cite{Daughton_etal-2009} 
will enable us to bridge the gap between two-fluid simulations, 
which implicitly assume some degree of collisionality, and purely 
collisionless PIC studies. These new capabilities will allow us 
to elucidate the transition between the slow collisional and fast 
collisionless reconnection regimes.

It is also important to invest resources into developing the analytical 
theory of reconnection. Indeed, most of the progress in the past decade
has been purely numerical and a complete theoretical understanding of 
reconnection is still lacking. This is important because most astrophysical 
systems where reconnection takes place are characterized by much greater 
separation of scales than can be achieved in simulations or in the lab.
Although a number of new and promising theoretical ideas, both on 
the fundamental issues of magnetic reconnection and on its astrophysical 
applications, have been proposed in recent years,
%(e.g., Lazarian \& Vishniac 1999, Spruit~et~al.~2001, 
%Kulsrud~2001, Rogers~et~al.~2001, Uzdensky 2003, 2007, 
%Malyshkin~et~al.~2005, Lyutikov \& Uzdensky 2003, 
%Lyutikov~2003, 2006, Lyubarsky 2005, Giannios \& Spruit 2005, 
%de Gouveia dal Pino \& Lazarian 2005, Uzdensky \& Kulsrud 2006, 
%Drake~et~al.~2006, Kulsrud 2005, Wang~et~al.~2008), 
this positive momentum will be lost unless continuously supported.

%---------------------------------------------------------------

On the experimental side, we advocate for the next generation,
medium-scale laboratory experiments aimed at addressing the 
above science questions.
In particular, a larger size is needed to have better diagnostics 
that would enable one to address the energy-partitioning problem, 
including non-thermal particle acceleration. In addition, a larger 
size will make it possible to reach sufficient scale separation 
to verify the formation of the so-called Petschek structure, which 
hasn't been seen in reconnection experiments so far. Also, a longer 
current sheet is needed to investigate the secondary tearing instability
in both collisional (resistive) and collisionless regimes, thus addressing
the bursty nature of magnetic reconnection and the associated particle 
acceleration (which is also one of the key directions of numerical research 
for the next few years). Finally, the new generation of experiments should 
enable studies of important global aspects of magnetic self-organization, 
e.g., the current-sheet formation process and the nonlinear dynamics of 
multiple current-sheet systems.

Another exciting experimental opportunity in the next decade 
is the availability of new powerful laser facilities. 
They will open up the high-energy density (HED) frontier in 
reconnection research, e.g., by enabling one to study the 
effects of radiation on HED reconnection, which will have 
important implications for high-energy astrophysics 
(e.g., magnetar flares).

%Recent studies of local reconnecting layer dynamics by 
%PIC codes, have learned that energy dissipation in the neutral sheet 
%occurs in a small region, leading to a much smaller rate of energy 
%conversion from magnetic to particle kinetic energy.  This rate is 
%too small to explain the observed particle heating during reconnection 
%observed in RFP plasma relaxation events, in spheromak merging experiments, 
%or in solar flare evolution.  

%As seen, for example, in solar flares and RFPs. 
%This point was made in 1995 by E. Lu who considered a simple automaton 
%model and demonstrated that if multiple reconnection layers were dynamically 
%related (such that one reconnection would trigger another), then one could 
%explain the singular result that the distribution function for the number 
%of flares of a given energy is a nearly perfect power law.  Finally, as 
%suggested by Parker in 1979, it would be of great value to develop and 
%elucidate a general theory of current layer formation in a highly 
%non-symmetric magnetic equilibrium such as is observed on the sun.  
%In RFP plasmas, reconnection in multiple layers are observed to generate 
%a significant of the global plasma invoking 
%strong ion heating, which is currently under intensive investigation.  
%A theory from a first principle may lead us to a breakthrough for solving 
%this problem. 

%-----------------------------------------------------------------

DAU wishes to thank A.~Bhattacharjee, E.~Blackman, S.~Boldyrev, W.~Daughton, 
J.~Goodman, A.~Lazarian, and E.~Zweibel for useful comments and suggestions.

%********************************************************************

\vskip15 pt
\begin{multicols}{2}

\renewcommand{\baselinestretch}{0.8} % Uncomment for 0.8 spacing between lines

\small

%\doublecolumn

% REFERENCES

\end{multicols}

%-----------------------------------------------------------------

%--------------------------------------------------------------

%--------------------------------------------------------------

\end{document}